\documentclass[fleqn,usenatbib]{mnras}

\usepackage{newtxtext,newtxmath}

\usepackage[T1]{fontenc}
\usepackage{ae,aecompl}

\usepackage{graphicx}	
\usepackage{amsmath}	
\usepackage{amssymb}	

\newcommand{\nneel}{$\approx 80$}  
\newcommand{\mdmdla}{0.25 {\rm pc \, cm^{-3}}} 
\newcommand{\dlali}{1\,{\rm AU}}

\newcommand{\aval}{0.236_{-0.021}^{+0.016}}
\newcommand{\bval}{0.168_{-0.017}^{+0.010}}
\newcommand{\cval}{2.87_{-0.13}^{+0.17}}

\newcommand{\nenHbp}{-2.881}
\newcommand{\nenHm}{0.352}  

\newcommand{\admtwo}{0.01}
\newcommand{\uppadm}{0.22}

\newcommand{\mdmunits}{{\rm pc \, cm^{-3}}} 
\newcommand{\dmunits}{$\mdmunits$}
\newcommand{\mnh}{n_{\rm H}}
\newcommand{\mnhi}{N_{\rm HI}}
\newcommand{\nhi}{$N_{\rm HI}$}
\def\cm#1{\, {\rm cm^{#1}}}
\def\N#1{N({\rm {#1}})}

\newcommand{\mlya}{{\rm Ly\alpha}}

\newcommand{\lya}{Ly$\alpha$}

\def\rtp{\, \right  ) }
\def\ltp{\left  ( \,}
\def\intl{\int\limits}

\newcommand{\mloz}{\ell_{\rm DLA}(z)}
\newcommand{\loz}{$\mloz$}
\newcommand{\mgoz}{g(z)}
\newcommand{\goz}{$\mgoz$}
\newcommand{\mfnhi}{f(\mnhi,z)}
\newcommand{\fnhi}{$\mfnhi$}
\newcommand{\mzdla}{z_{\rm DLA}}
\newcommand{\zdla}{$\mzdla$}
\newcommand{\mxdla}{x_{\rm DLA}}
\newcommand{\xdla}{$\mxdla$}

\newcommand{\mdmnm}{{\rm DM}^{\rm NM}_{\rm DLA}}
\newcommand{\dmnm}{$\mdmnm$}
\newcommand{\mdmwim}{{\rm DM}^{\rm WIM}_{\rm DLA}}

\newcommand{\mdmaliii}{{\rm DM}_{\rm AlIII}}
\newcommand{\dmaliii}{$\mdmaliii$}
\newcommand{\mdmodla}{{\rm DM}_{\rm DLA}}
\newcommand{\dmodla}{$\mdmodla$}
\def\mfavgdm#1{\overline{DM}^{\rm NM}_{\rm DLA}(\mzfrb={#1})}
\newcommand{\mavgdm}{\overline{DM}^{\rm NM}_{\rm DLA}(\mzfrb)}
\newcommand{\avgdm}{$\mavgdm$}
\newcommand{\mavgn}{\bar{n}_{\rm DLA}(z_{\rm FRB})}
\newcommand{\avgn}{$\mavgn$}
\newcommand{\mzfrb}{z_{\rm FRB}}
\newcommand{\zfrb}{$\mzfrb$}
\newcommand{\mascatt}{\theta_{\rm scatt}}
\newcommand{\ascatt}{$\mascatt$}
\newcommand{\mtscatt}{\tau_{\rm scatt}}
\newcommand{\tscatt}{$\mtscatt$}
\newcommand{\mrdiff}{r_{\rm diff}}
\newcommand{\rdiff}{$\mrdiff$}
\newcommand{\msmeff}{{\rm SM}_{\rm eff}}
\newcommand{\smeff}{$\msmeff$}

\newcommand{\mdlatau}{\tau_{\rm DLA}}
\newcommand{\dlatau}{$\mdlatau$}
\newcommand{\mtmin}{\tau_{\rm min}}
\newcommand{\tmin}{$\mtmin$}

\title[DLAs and FRBs]{The Astrophysical 
Consequences of Intervening Galaxy Gas
on Fast Radio Bursts}

\author[Prochaska \& Neeleman]{
J. Xavier Prochaska,$^{1}$\thanks{E-mail: xavier@ucolick.org}
Marcel Neeleman,$^{1}$
\\
$^{1}$Astronomy \& Astrophysics, UC Santa Cruz, 1156 High St., Santa Cruz, CA 95064 USA
}

\date{Accepted XXX. Received YYY; in original form ZZZ}

\pubyear{2017}

\begin{document}
\label{firstpage}
\pagerange{\pageref{firstpage}--\pageref{lastpage}}
\maketitle

\begin{abstract}
We adopt and analyze results on the incidence and
physical properties of damped \lya\ systems (DLAs) to predict
the astrophysical impact of gas in galaxies on observations
of Fast Radio Bursts (FRBs).  
Three DLA measures form the basis of this analysis:
(i) the \ion{H}{I} column density distribution, 
parameterized as a double power-law;
(ii) the incidence of DLAs with redshift (derived here),
$\mloz=A+B \,\arctan(z-C)$ with 
$A=\aval,B=\bval,C=\cval$;
and
(iii) the electron density, parameterized
as a log-normal deviate with mean
$10^{-2.6}\cm{-3}$ and dispersion 0.3\,dex.
Synthesizing these results, we estimate that the average 
rest-frame dispersion measure 
from the neutral medium of a single, intersecting
galaxy is $\mdmnm=\mdmdla$. 
Analysis of \ion{Al}{III} and \ion{C}{II}* absorption
limits the putative  warm ionized medium
to contribute $\mdmwim<20\mdmunits$.
Given the low incidence of DLAs, we find that
a population of FRBs at $z=2$ will incur
$\mfavgdm{2}=\admtwo\mdmunits$ on average, 
with a 99\%\ c.l.\ upper bound 
of $\uppadm\mdmunits$.
Assuming that turbulence of the ISM in external galaxies
is qualitatively similar to our Galaxy, we estimate that
the angular broadening of an FRB by intersecting galaxies 
is negligible ($\mascatt<0.1$\,mas).  
The temporal broadening is also predicted to be small, 
$\mdlatau\approx 0.3\;\rm ms$ for a $z=1$ galaxy
intersecting a $z=2$ FRB for an observing frequency
of $\nu=1$\,GHz.
Even with $\nu=600$\,MHz, the fraction of sightlines
broadened beyond 25\,ms is only approximately 0.1\%.  
We conclude that gas within the ISM of
intervening galaxies has a minor effect on the detection 
of FRBs and their resultant DM distributions.
Download the repository at https://github.com/FRBs/FRB
to repeat and extend the calculations  presented here.
\end{abstract}




\section{Introduction}

The discovery of the `repeating' Fast Radio Burst \citep[FRB]{frb121102}
and subsequent follow-up observations at the Very Large Array \citep{chatterjee17}
have lead to the confirmation that at least some FRB events have
an extragalactic origin \citep{tendulkar17}.
In turn, the large Dispersion Measure (DM) that essentially defines
an FRB offers one the opportunity to study the integrated electron
column density along individual sightlines through the universe.
New and upcoming surveys -- e.g.
CHIME \citep{CHIME}, 
ASKAP \citep{bannister17a}, 
APERTIF, 
REALFAST \citep{realfast16} --
will yield a terrific set of observations to measure
baryons in the $z<1$ universe and possibly
beyond, complementing decades of work in the far-UV
\citep[e.g.][]{dt01,cpw+05,tejos+14}.

For a source at great distance, e.g.\ $z \sim 1$ or approximately
2.4~Gpc, one expects a significant DM from the diffuse, highly-ionized
intergalactic medium \citep[IGM;]{inoue04}.  In quasar absorption
line (QAL) parlance, this plasma is referred to as the \lya\ forest.
Gas in the dark matter halos of galaxies will also
contribute and may even dominate 
\citep[; Prochaska \& Zheng 2017, in prep.]{mcquinn14}.
This gas in QAL research is referred to
as the circumgalactic  medium (CGM), 
which one frequently associates to the optically thick
Lyman limit systems \citep[LLSs; e.g.][]{fumagalli11a,ribaudo11,hafen16}.
Most rarely, an FRB sightline may 
penetrate the ISM of a galaxy akin to our own.  If the neutral hydrogen
column density \nhi\ equals or exceeds
$2\times 10^{20} \cm{-2}$, the QAL community
refers to the absorption as a damped \lya\ (DLA) system 
\citep{wolfe86,wgp05}.  
In the far-UV, a DLA system shows a quantum 
mechanically `damped' \lya\ line with equivalent width
$W_\mlya > 10$\AA.
In addition to contributing to the DM value of an FRB,
the free electrons in DLAs -- if turbulent -- 
will scatter the radio pulse \citep{macquart13}.
Turbulent scattering broadens both the angular size of the source and 
the intrinsic duration of the event.

The DLA contribution to DM
is distinct from other intervening gas because:
  (1) DLAs arise in collapsed (i.e.\ highly non-linear) structures; and
  (2) they trace a predominantly neutral gas with ionization
  fraction $x \equiv n_{\rm H^+}/\mnh \ll 1$. 
We emphasize that the analysis performed here considers primarily
the neutral gas of DLAs, i.e.\ the warm and cold neutral media
(WNM/CNM) of external galaxies.  We also consider
a putative warm ionized medium (WIM) surrounding DLAs.
We explicitly ignore, however, gas in the dark
matter halos hosting DLAs which is believed to be traced
by high ions like C$^{+3}$ and Si$^{+3}$ 
\citep{wp00a,wp00b,mps+03,fpl+07,pcw+08}.
The astrophysical impact of galactic halos on FRBs
will be treated in a companion paper 
(Prochaska \& Zheng 2017, in prep.; see also McQuinn 2014).

We recognize that the DLA criterion is somewhat arbitrary
and gas with modestly lower column densities
(the so-called super Lyman limit systems or sub-DLAs)
may be qualitatively similar and may also arise in the
ISM of intervening galxies. However, most of these systems trace
gas in the halos of galaxies and therefore we will 
examine such systems explicitly in the companion paper. We note that
the code provided with this paper can easily be modified to 
include such material within the framework developed here.

It has been proposed that intervening galaxies will essentially 
`extinguish' the FRB signal by broadening the pulse duration
to a non-detectable signal  \citep{macquart13,mcquinn14}.
Analogous to dust obscuring UV-bright quasars to bias observed 
sightlines against a subset of DLAs \citep{oh84,fall93},
it is possible that the observed FRB population will not occur along
sightlines that intersect galaxies.  In turn, this may bias the observed
DM distribution and alter the observed
redshift distribution of FRBs. 
A precise characterization of these effects is therefore critical
to inferring the intrinsic nature of FRBs and for utilizing 
the population to constrain 
cosmological quantities \citep[e.g.][]{gaensler08}.

In this paper, we analyze the latest results from DLA surveys
to empirically estimate the impact of DLAs on current and future
FRB studies.  This includes statistics on the incidence of DLA
absorption \loz, the frequency distribution of \nhi, 
aka \fnhi, and constraints on the electron volume density
$n_e$.  Throughout, we adopt the Planck 2015 cosmology
\citep{Planck2015} encoded in {\it astropy} (v1.3).
All of the calculations presented here may be reproduced
and extended using codes in this repository:
https://github.com/FRBs/FRB.

\section{The Dispersion Measure of DLAs}

As emphasized above, gas in intervening galaxies will contribute
to the DM of FRBs for those sightlines that intersect
any such galaxies.  On the reasonable ansatz that {\it all}
galaxies with a non-negligible DM will also have an \nhi\ value
satisfying the DLA criterion, we may use statistics and measured
properties of the latter to predict the effects of the former.
This forms the basis of our methodology.

We may construct an empirical estimate for the average DM value
from the neutral medium in DLAs for a FRB 
at redshift \zfrb, \avgdm, as follows. Formally, we have,

\begin{equation}
\mavgdm = \intl_0^{z_{\rm FRB}} \intl_{10^{20.3}}^\infty 
  \mfnhi \, \mnhi \, \mxdla(\mnhi) (1+z)^{-1} \; d\mnhi \, dz \;\;\; ,
\end{equation}
where we have defined $\mxdla \equiv n_e/n_{\rm H}$ with
$n_{\rm H} \approx n_{\rm HI}$ for the predominantly neutral DLAs.
One recognizes \fnhi\ as the \ion{H}{I} frequency distribution, i.e.\
$\mfnhi \, d \mnhi \, dz$ is 
the estimated number of DLAs in \nhi\ and redshift intervals.

Thus far, DLA surveys have not revealed a significant evolution in 
the shape of \fnhi\ with redshift\footnote{Note that this 
differs from the full population of QALs, e.g.\ \cite{wp11}.}.
Therefore, we will assume that the function describing
\fnhi\ is separable \citep[see also][]{inoue+14}, i.e.

\begin{equation}
\mfnhi = h(\mnhi) \, j(z) \;\;\; .
\end{equation}
For the first term, we adopt the double power-law derived by 
\cite{phw05} from their analysis of the SDSS-DR5:

\begin{equation}
h(\mnhi) =  K \ltp \frac{N}{N_d} \rtp^\beta  \; {\rm where} \; \beta =  
\begin{cases}
\alpha_3:  N < N_d ; \quad \\ 
\alpha_4:  N \geq N_d \\
\end{cases}
\label{eqn:dbl}
\end{equation}
with $N_d = 10^{21.551} \cm{-2}$, $\alpha_3 = -2.055$, and
$\alpha_4 = -6$.
While there have been more recent and larger surveys of DLAs
from BOSS \citep[e.g.][]{noterdaeme+12,bird17}, most 
have not provided functional fits to their \fnhi\ analysis
\footnote{An analysis of the complete SDSS survey yields
\fnhi\ with a fully consistent shape as the one adopted
here \citep{np+09}.}.  
Furthermore, our own efforts with deep-learning techniques
raises concerns on several of the reported results 
\cite{parks+17}.
In any case, as progress on DLA surveys continues,
we will ingest and update the accompanying repository\footnote{
https://github.com/FRBs/FRB} and the results on FRBs are
relatively insensitive to the precise form of $h(\mnhi)$.

\begin{figure}
	\includegraphics[width=\columnwidth]{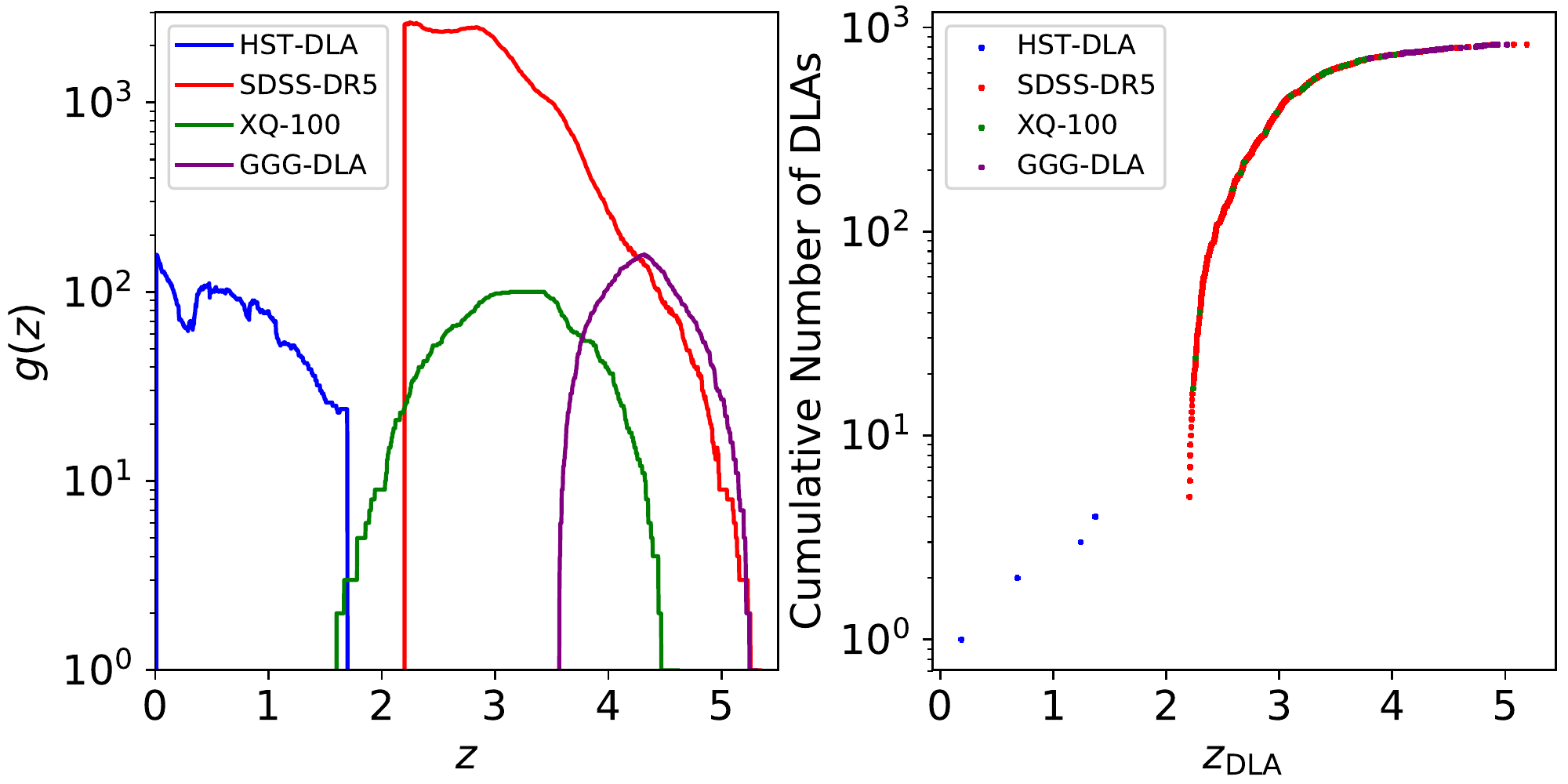}
    \caption{(left) The curves describe the survey path \goz\ 
    for the datasets used to estimate the
    DLA incidence with redshift.  The \goz\ curves indicate
    the number of quasars searched for DLAs at a given redshift.
    (right) The cumulative number of DLAs identified in these
    surveys, ranked by redshift.  Note the scarcity of DLAs
    at $z<2$ indicating a low incidence.
	}
    \label{fig:goz}
\end{figure}

To constrain $j(z)$,  we perform a new calculation
of the incidence of DLAs:

\begin{equation}
\mloz = \intl_{10^{20.3}}^\infty \mfnhi d\mnhi  
=  j(z) \intl_{10^{20.3}}^\infty h(\mnhi) d\mnhi  
\;\;\; .
\end{equation}
In the following, we set the normalization $K$ of $h(\mnhi)$
such that $\mloz = j(z)$, i.e. the integral over $h(\mnhi)$ is unity.
Many DLA surveys have been analyzed to assess \loz\ across cosmic
time \citep[e.g.][]{wlf+95,sw00,phw05,rao17}.
The standard approach is to first evaluate the search path
for DLAs, a.k.a. $g(z)$ or the number of independent quasar sightlines
surveyed for DLA absorption at a given redshift.
A good estimator for the incidence in a finite redshift
interval $\Delta z$ at redshift $z$ is 

\begin{equation}
\mloz = N_{\rm DLA} / \Sigma g(z) 
\label{eqn:lz}
\end{equation}
with $N_{\rm DLA}$ the number of DLAs detected in  
$\Delta z$, and the sum on $g(z)$
is performed over the same interval. 

The left panel of
Figure~\ref{fig:goz} summarizes the redshift path analyzed by
the DLA surveys adopted here: 
a blind search for DLAs at $z<2$ in Hubble Space Telescope
UV spectroscopy, HST-DLA \citep{neeleman+16};
a survey of quasars from the SDSS data release 5, 
SDSS-DR5 \citep{pw09};
a survey of 100 quasars at $z=3-4$ observed with the
X-Shooter spectrograph, XQ-100 \citep{xq_100,sanchez+16};
a survey for $z>3.5$ DLAs using the Gemini GMOS spectrometers,
GGG-DLA \citep{worseck+14,crighton+15dla}.
The cumulative distribution of DLAs detected, $N_{\rm DLA}$,
is shown in the right panel.
One point is obvious from the figure: despite the survey of 
many hundreds of quasars at $z<1$ with {\it HST},
only a handful of DLAs were detected\footnote{We do not include
here the DLA surveys of Rao \& Turnshek who targeted strong 
\ion{Mg}{II} absorbers to detect DLAs \citep{rtn06,rao17}.}.
This primarily follows from cosmological expansion, 
i.e. one predicts 
that the observed incidence decreases as
$(1+z)^2$  for an unevolving population of absorbers per 
comoving volume in a $\Lambda$-dominated universe. 

\begin{figure}
	\includegraphics[width=\columnwidth]{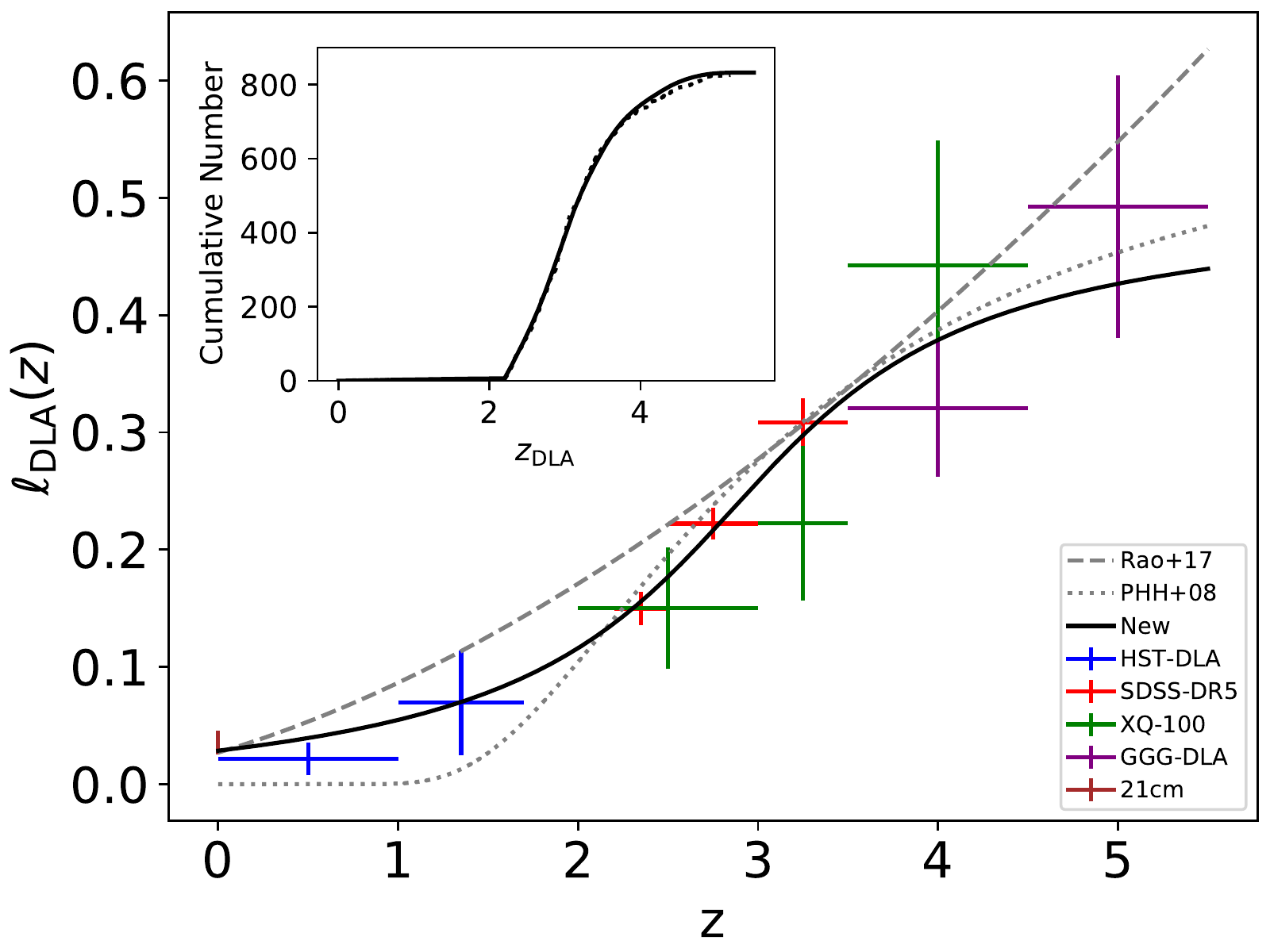}
    \caption{Incidence of DLAs per unit redshift \loz\ estimated
    from the surveys described by Figure~\ref{fig:goz}.  Overlayed
    on the binned evaluations (the $x$ bars show bin size and 
    the $y$ bars indicate Poisson uncertainty)
    are previous fits from the literature
    \citep[grey curves;][]{phh08,rao17} and our new derivation
    (black curve).  The inset compares the observed cumulative
    number of DLAs to the model, and a one-sided Kolmogorov-Smirnov
    test yields an acceptable probability for the null hypothesis.
    }
    \label{fig:loz}
\end{figure}

From the data in Figure~\ref{fig:goz}, we have used the 
estimator in Equation~\ref{eqn:lz} to calculate \loz\ in
select bins (Figure~\ref{fig:loz}).  Here we derive 
uncertainties assuming Poisson statistics.  It is apparent that \loz\
increases with redshift from low values at 
$z<1$.  The figure also show an estimate for $\ell(z)$ 
at $z \approx 0$ based on 21\,cm observations 
where we have combined the values
reported by \cite{zms+05} and \cite{braun12}:
$\ell(z)_{\rm 21 \, cm} = 0.035 \pm 0.01$.

Most previous analyses of \loz\ have assumed it follows the 
functional form $(1+z)^\gamma$. This is physically motivated by cosmological
expansion, but \cite{phh08} emphasized that the evolution
in \loz\ at $z \approx 2-3$ is not well described by such a power-law.
Lacking a proper physical model for \loz, we take an empirically 
driven approach and seek a functional form that describes the
data well at all redshifts and with the fewest number of parameters.
After some experimentation, we settled on a model,

\begin{equation}
\mloz = A + B \, \arctan(z-C) \;\;\; ,
\end{equation}
which captures an apparent inflection in \loz\ at the highest 
redshifts studied \citep{crighton+15dla}.
We performed a standard maximum likelihood
analysis to find the best values for $A,B$ and $C$ as shown
in Figure~\ref{fig:loz} and listed in Table~\ref{tab:param}.
Errors on these parameters were estimated using standard
bootstrap techniques. 
All of the data and our new \loz\ result are ingested in the 
{\it pyigm}\footnote{https://github.com/pyigm/pyigm} package.

\begin{figure}
	\includegraphics[width=\columnwidth]{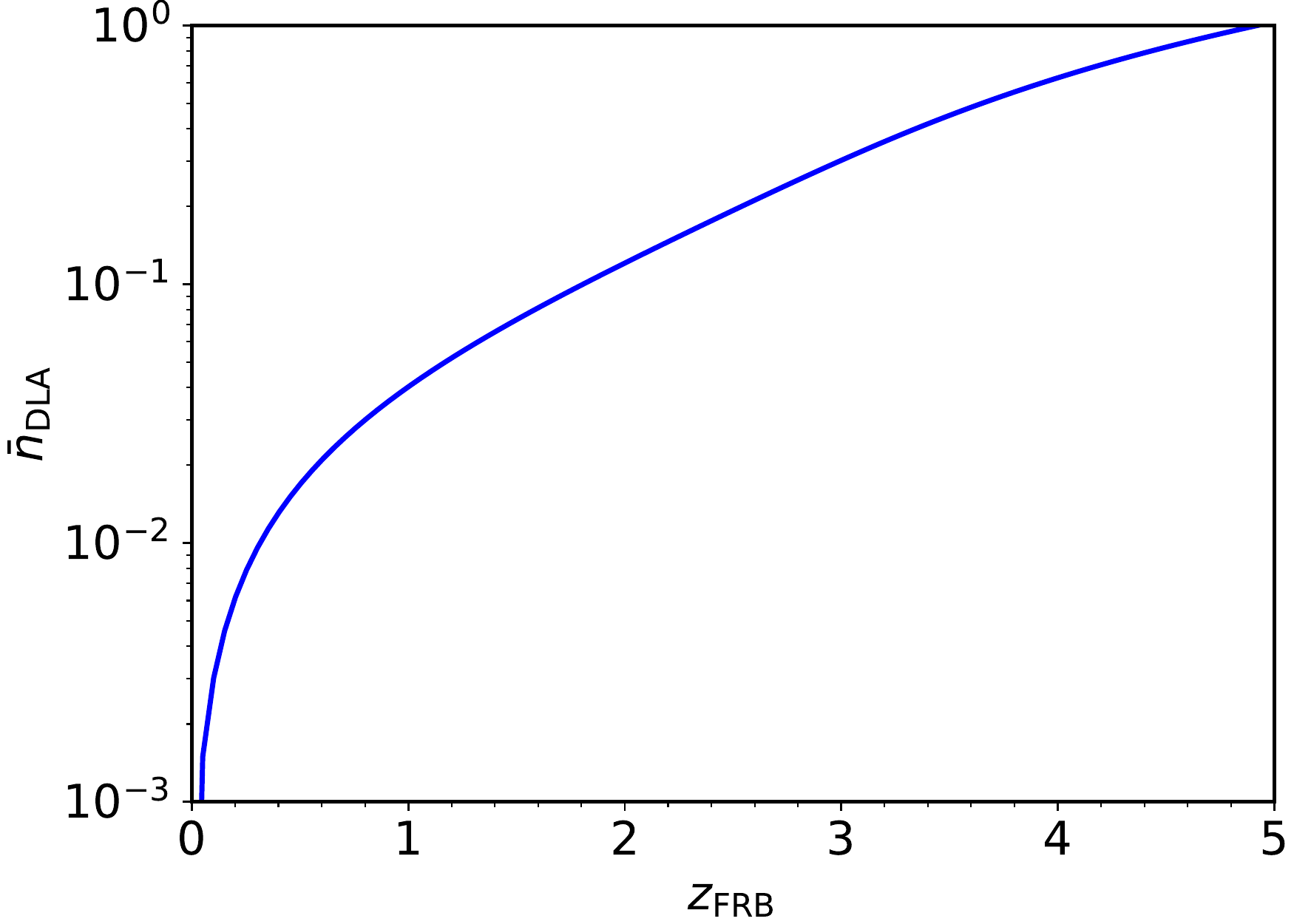}
    \caption{Average number of DLAs \avgn\ intersected by a sightline
    to a source at redshift \zfrb.  For $\mzfrb=1$, we estimate
    $\mavgn < 0.05$ and the value remains less than unity until
    $\mzfrb \approx 5$.  For a population of events at a given 
    redshift, the number of DLAs observed will be Poisson distributed
    with mean and variance of \avgn.
	This relatively low incidence strictly limits the potential
	impact of intervening galaxies on FRBs.
    }
    \label{fig:avgn}
\end{figure}

With \loz\ well-fitted, we may calculate the average number of DLAs
intervening a source at redshift $z_{\rm FRB}$ as:

\begin{equation}
\begin{aligned}
\mavgn &= \intl_0^{\mzfrb} \, \mloz \, dz \\
&= A \, z_{\rm FRB} + B \, [\mu \arctan\mu - \ln(\mu^2 + 1)/2 \\
- C \arctan C + \ln(C^2 + 1)/2 ]
\end{aligned}
\label{eqn:nz}
\end{equation}
with $\mu \equiv C-\mzfrb$.
For a single sightline to an FRB, the uncertainty in \avgn\ is dominated
by Poisson sampling with $\sigma^2(\bar n_{\rm DLA}) = \bar n_{\rm DLA}$.
For a population of FRBs, the uncertainty is dominated 
by errors in our estimation of \loz.  
Figure~\ref{fig:avgn} shows \avgn\ versus redshift, where one
recovers a value of 0.04 for $\mzfrb = 1$ and 
one notes \avgn\ remains less than unity until nearly $\mzfrb = 5$.
This relatively low incidence strictly limits the potential
impact of intervening galaxies on FRBs.

Owing to the large \ion{H}{I} column densities that define DLAs,
the gas is (extremely) optically thick to \ion{H}{I} ionizing
radiation and one expects DLAs to be predominantly neutral gas.
This has been demonstrated theoretically and empirically 
\citep[e.g.][]{viegas95,vladilo01}, although there are notable,
individual counter-examples \citep[e.g.][]{pro_ion02}.
A direct assessment of the neutral fraction of DLAs (or any QAL)
is challenged by the fact that the commonly observed resonance
lines are insensitive to the physical conditions of density,
temperature, etc.
Therefore, direct measurements of \xdla\ are difficult to obtain.  

\begin{figure}
	\includegraphics[width=\columnwidth]{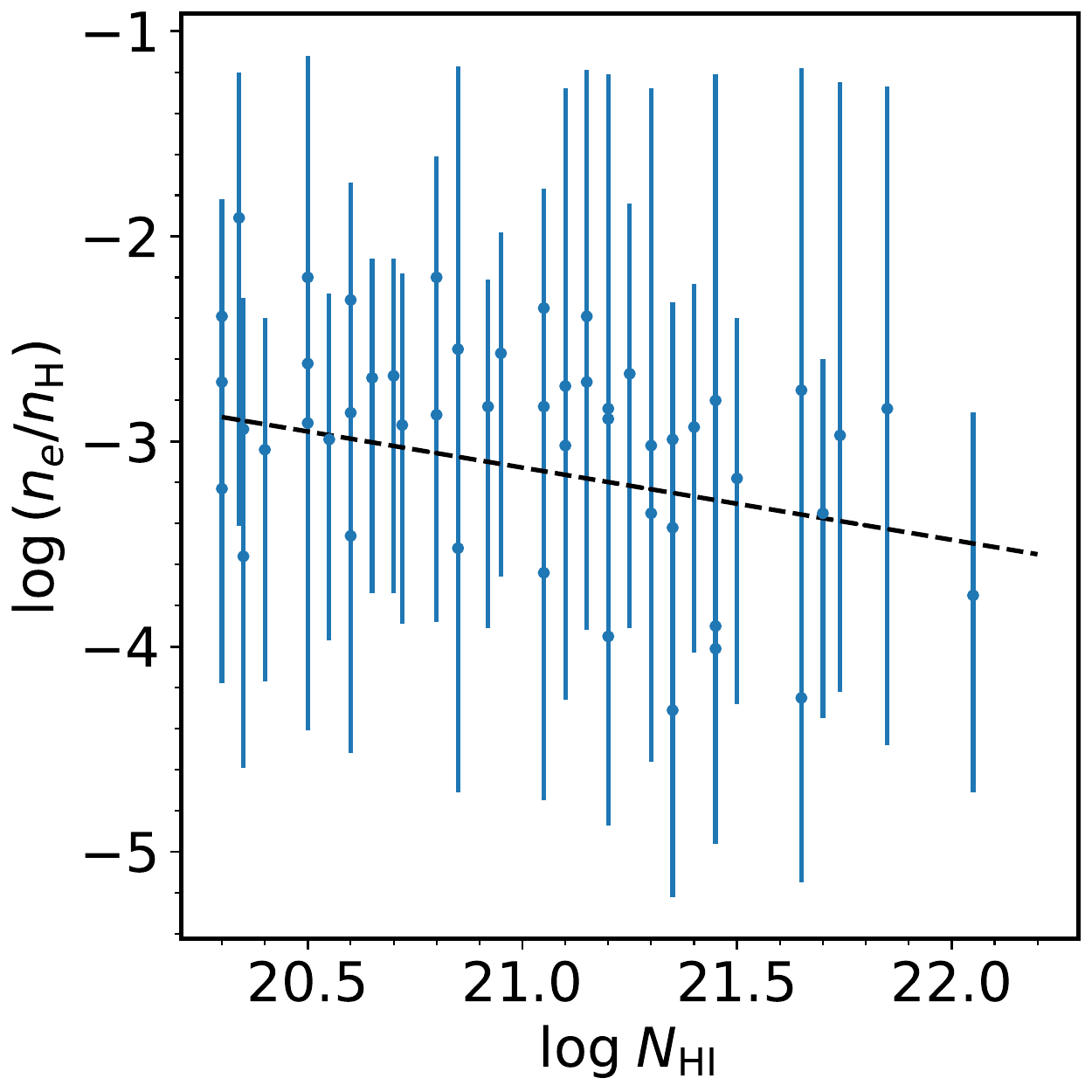}
    \caption{Estimates of the ionization fraction of DLA gas
    ($\log x_{\rm DLA} = \log[n_e/n_{\rm H}])$
    from an MCMC analysis of measured column densities of 
    ionized carbon and silicon and their associated excited
    levels \citep{neeleman+15}.  Overplotted on their sample
    is a log-linear fit, described by Equation~\ref{eqn:nenH}.
    }
    \label{fig:nenH}
\end{figure}

A notable exception is to assess the physical conditions
within DLAs by analyzing
absorption from the excited states of C$^+$ and Si$^+$
\citep{howk05}.
The most comprehensive study to date was by 
\cite{neeleman+15} who
used an MCMC analysis to derive estimates of the electron
and hydrogen number densities $n_e$, $\mnh$ for \nneel\ DLAs at 
$z \approx 2-5$.  
Figure~\ref{fig:nenH} summarizes the results versus the
\ion{H}{I} column density for the 50 systems with \ion{C}{II}* detected.  
%
There is a small trend of decreasing ionization fraction 
with \nhi\ that we parameterize as, 

\begin{equation}
\log \, \mxdla = \log \frac{n_e}{\mnh} = \nenHbp - \nenHm \, (\log \mnhi - 20.3) \;\;\; .
\label{eqn:nenH}
\end{equation}
We proceed by  assuming an 0.5\,dex uncertainty in $n_e$
for any given DLA.  

We now have all the ingredients necessary to offer an 
estimate for the average DM for a sightline intersecting
one galaxy, 

\begin{equation}
\mdmnm = \frac{\intl_{10^{20.3}}^\infty h(\mnhi) \mnhi \mxdla \, d\mnhi}{
   \intl_{10^{20.3}}^\infty h(\mnhi) d\mnhi} \approx <\mnhi> \mxdla(<\mnhi>) 
   = \mdmdla
\label{eqn:DM}
\end{equation}
This is substantially smaller than the lowest values measured
for sightlines intersecting our own ISM \citep[$\approx 25$\dmunits,
from the Sun]{gaensler08}.

We restrict our analysis to \ion{C}{II}* detections because
non-detections yield little constraint on the electron density.
However, by restricting the $n_e/n_{\rm H}$ assessment to DLAs 
with \ion{C}{II}* detections, we may be biased towards sightlines
dominated by the CNM \citep{wpg03,neeleman+15}, and therefore lower
ionization fractions. Indeed, models of a multi-phase ISM
predict ionization fractions for the WNM phase of $x_{\rm WNM} \approx 0.1$
\cite[e.g.][]{wolfire95}. We note, however, that DLAs without 
\ion{C}{II}* detections have systematically lower \nhi\ values. 
For the \dmnm\ measurement, the ionization fraction gets weighted 
by \nhi\ and the bias resulting from omitting these systems is 
thereby at least partially offset. An additional bias toward 
sightlines dominated by the CNM could have been introduced from the 
selection criteria adopted by \citet{neeleman+15}. However, even
if we assume that a majority of the gas arises in the WNM, as is
indicated by 21\,cm observations of radio-loud quasars with intervening
DLAs \citep{kc03,kanekar+14}, the primary conclusions of the paper
remain unchanged.

\begin{figure}
	\includegraphics[width=\columnwidth]{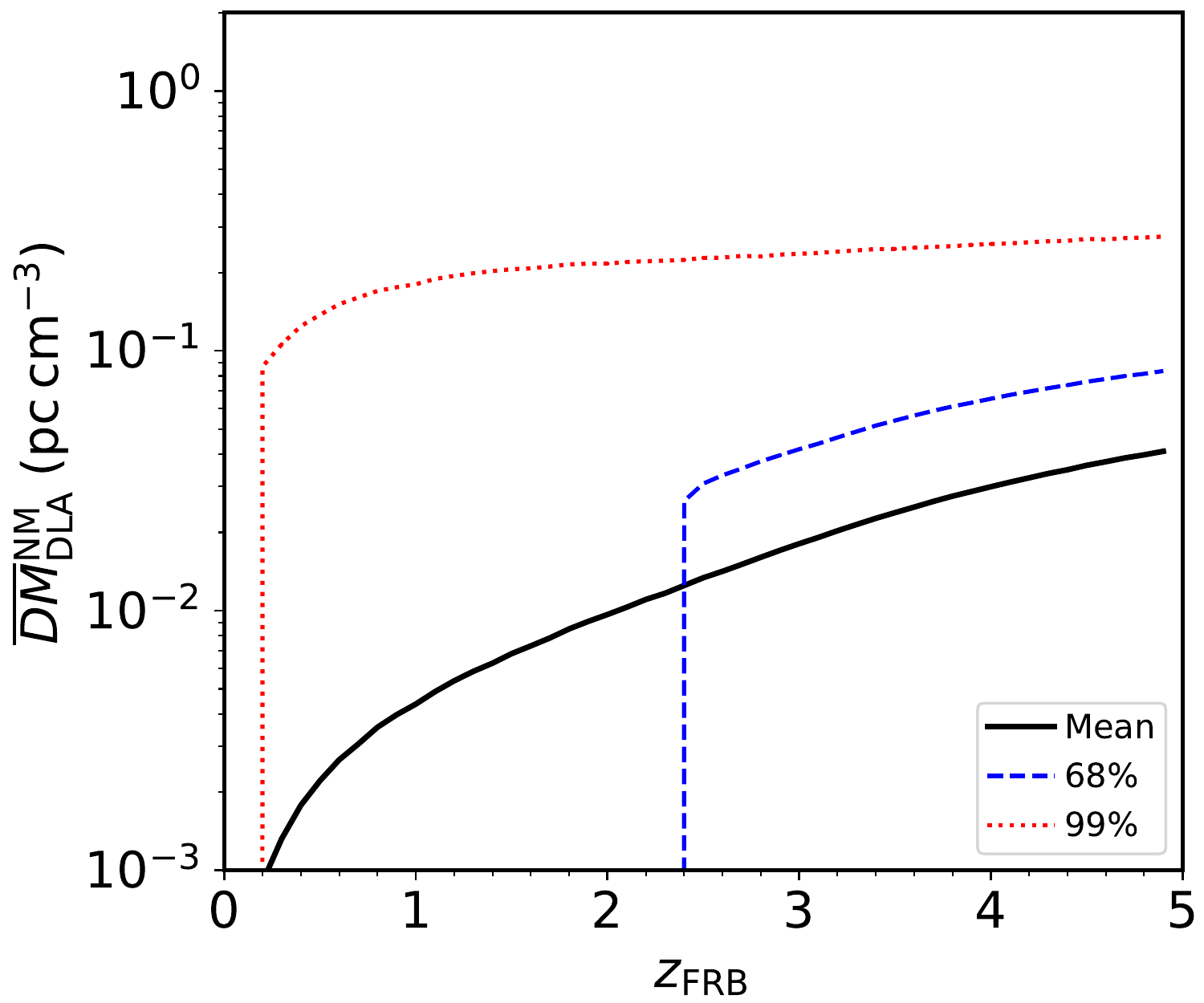}
    \caption{The solid, black curve shows the average DM due
    to DLAs measured from  a population of FRBs at redshift 
    \zfrb.  For the CNM/WNM of these galaxies, we find
    $\mavgdm < 10^{-1} \mdmunits$ at all $\mzfrb < 7$.
    The blue-dashed and red-dotted curves show the 68\%\ and 99\%\
    upper intervals for \avgdm\ calculated from $10^6$ random
    draws.  These equal zero at low \zfrb\ because of the 
    low probability to intersect even a single galaxy
    (Figure~\ref{fig:avgn}).
    }
    \label{fig:avgDM}
\end{figure}

Because the shape of \fnhi\ is taken to be invariant with 
redshift, the $d\mnhi$ integral in Equation~\ref{eqn:DM}
may be evaluated independently
to derive an approximate expression for the average contribution
for a population of FRBs at \zfrb:

\begin{equation}
\mavgdm \approx \mdmodla \, \mavgn \, (1+\bar z)^{-1}
\label{eqn:approxDM}
\end{equation}
with $\bar z$ the average redshift of intervening galaxies.
Evaluating Equation~\ref{eqn:approxDM} at $\mzfrb = 1$, 
we estimate 
$\overline{DM}^{\rm NM}_{\rm DLA}(\mzfrb=1) \approx 0.0065 \mdmunits$.
To generate a more accurate evaluation of \avgdm, we have
performed a Monte Carlo simulation of $10^6$ sightlines
to FRBs over a series of \zfrb\ values.
For each realization, we draw the number and redshifts
of DLAs on the sightline according to \loz, 
(allowing for Poisson variance in \avgn), assign random
\nhi\ values drawn from $h(\mnhi$), 
evaluate $x_{\rm DLA}$, and calculate the DM value.
The results are shown in 
Figure~\ref{fig:avgDM}, and we find small
values at all \zfrb.
We also show the
68\%\ and 99\%\ intervals and find
that $\sim 1\%$ of FRB sightlines originating
at $\mzfrb = 5$ will experience a contribution of $\sim 0.2 \mdmunits$
from the neutral ISM of intervening galaxies.

As emphasized above, in our Galaxy the smallest DM values
measured
from our position at the Sun are approximately 
25\dmunits.  
It is accepted that a warm ionized medium (WIM),
which also generates nearly ubiquitous H$\alpha$ emission
across the sky \citep[e.g.][]{rth+98},
is responsible for this `floor' to the Galactic DM
\citep[e.g.][]{ne2001}.  
Several studies have used the observed DM distribution 
of pulsars with Galactic latitude and distance
to model the density,
filling factor, and distribution with scale height
of electrons in the WIM \citep{gaensler08,bf08}.

The Galactic WIM is believed to be primarily generated by photons
from O stars that have `leaked' through their \ion{H}{II}
regions \citep{haffner09}.  
Many external galaxies at $z \sim 0$ also exhibit a WIM
component typically referred to as diffuse ionized gas or DIG
\citep[e.g.][]{rand97}.
At high $z$, the presence and nature of the WIM in galaxies
has not yet been established.  One notes that the average 
star formation rate in high-$z$ galaxies is higher 
which could accentuate a WIM, but the ISM may also be
denser and less porous.  Separately, the
extragalactic UV background, which is more intense
at high $z$, may ionize the outer ISM of galaxies to
produce a WIM component.


In analogy with the Galactic ISM, it is possible that
DLAs are also enveloped in layers of predominantly
ionized gas, i.e.\ a WIM component\footnote{
Note again that we distinguish here from gas in the dark matter
halo presumed to contain DLAs.  This CGM component is discussed
in a separate manuscript (Prochaska \& Zheng 2017, in prep.; 
see also McQuinn 2014). }.
Indeed, \cite{howk_ion99} first emphasized that the frequent
detection of Al$^{++}$ in DLAs suggests 
a putative warm ionized medium, noting that \ion{Al}{III}
absorption in our Galaxy traces its underlying
electron distribution \citep{savage90}.
Furthermore, the \ion{Al}{III} absorption in DLAs more frequently
traces the low-ion transitions instead of the more highly ionized
gas revealed by e.g.\ \ion{C}{IV} \citep{wp00a}.
On the other hand, one may ionize a non-negligible fraction of
Al$^+$ to Al$^{++}$ via hard radiation from local sources
or the EUVB without significantly ionizing
hydrogen because of the large cross-section of Al$^+$ to 
high-energy photons \citep{sj98,pro_ion02}.  In any event, we
estimate an additional contribution to DM from DLAs based on their
$\N{Al^{++}}$ measurements.   This estimate will also
include any contribution from a WNM component that was
not captured by our treatment above.

\begin{figure}
	\includegraphics[width=\columnwidth]{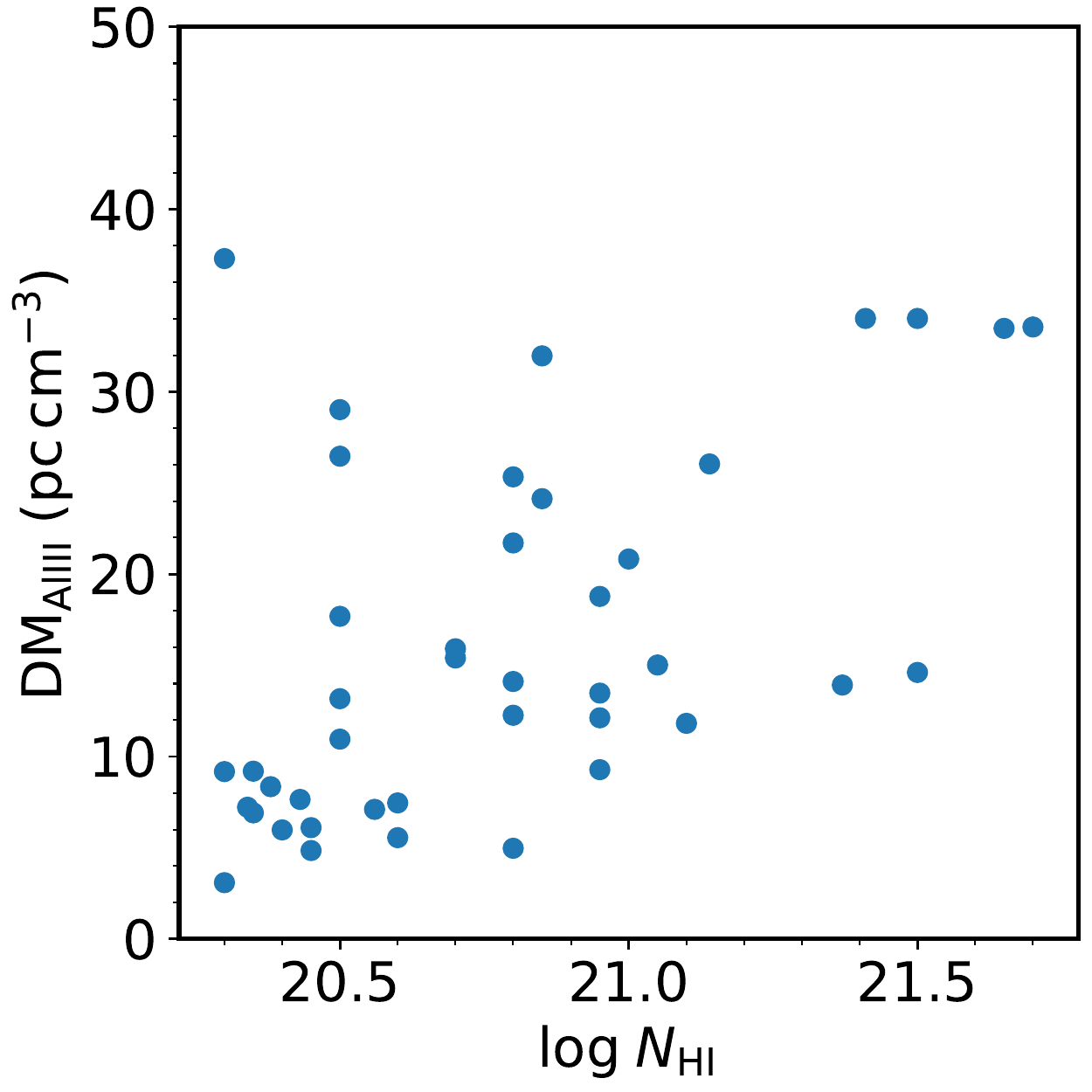}
    \caption{The points track estimates of DM for a putative WIM
    component in DLAs by adopting Equation~\ref{eqn:AlIII} and 
    the Al$^{++}$ column density measurements and metallicities
    reported for $z \sim 2$  (mention pyigm).
    For sightlines intersecting fewer than $10^{21}$
    hydrogen atoms, we infer $\mdmaliii < 30 \mdmunits$.
    The positive correlation between \dmaliii\ and \nhi\
    implies that a significant fraction of the observed
    \ion{Al}{III} absorption arises in neutral gas and,
    therefore, that \dmaliii\ likely overestimates the
    true contribution to DM from any WIM component.
    }
    \label{fig:AlIII}
\end{figure}

Specifically, we assume
that Al$^{++}$ traces a predominantly ionized gas
($\mxdla \approx 1$) and that it is the dominant ionization
state of Al in this plasma.  It follows that,

\begin{equation}
N_e \approx \N{Al^{++}} - \epsilon({\rm Al}) + 12.  
- \log(Z/Z_\odot) \;\;\; ,
\label{eqn:AlIII}
\end{equation}
with $\epsilon({\rm Al})$ the number abundance of Al
in the Sun \citep[6.43;][]{asplund09} and $Z/Z_\odot$
the metallicity of the DLA relative to solar.
Figure~\ref{fig:AlIII} shows \dmaliii\ values derived with
Equation~\ref{eqn:AlIII} versus \nhi\
using the HIRES sample of \cite{marcel13};
statistical uncertainties of approximately 0.2\,dex
are dominated by the error in $Z/Z_\odot$.
Even with the somewhat extreme assumptions encoded by 
Equation~\ref{eqn:AlIII}, we derive a median
\dmaliii\ value of less than $20 \mdmunits$.
One also notes that the highest \dmaliii\ values are
associated primarily with high \nhi\ DLAs suggesting 
that this gas arises in the WNM/CNM. 
Altogether, we constrain
$DM_{\rm DLA}^{\rm WIM}$ to be less 
than 20\,pc\,cm$^{-3}$ and likely less
than 10\,pc\,cm$^{-3}$. 
This upper limit is also consistent with the many 
non-detections of \ion{C}{II}* absorption
in DLAs \citep{wpg03}.

\section{Scattering}
\label{sec:scatt}

The radio emission from a distant, compact source that intersects
a DLA may be scattered by electron
density inhomogeneities in the gas.
The effects include an angular broadening of the source
\ascatt\ and temporal broadening \tscatt.  Both
exhibit a frequency dependence.
The scattering processes has been studied extensively for
sightlines through the ISM to
pulsars in our Galaxy \citep{tc93}, and 
the formalism has been developed for gas intersecting
FRBs by \cite[][; hereafter M13]{macquart13}.  
We do not repeat their
complete derivation here but present key expressions,
treating angular and temporal broadening separately.

\subsection{Angular Broadening}
\label{sec:ascatt}

The angular broadening of an image by a point source has 
radius (half-width at half-maximum),

\begin{equation}
\mascatt = f \frac{D_{LS}}{D_{S} k \mrdiff} \;\;\; ,
\end{equation}
with $f$ a factor of order unity, $D_{LS}$
and $D_S$ the angular diameter distances 
from the source to the phase plane and to
the source respectively, $k \equiv 2 \pi / \lambda_0$
with $\lambda_0$ the wavelength in the observer 
frame, and \rdiff\ parameterizes the phase structure
function in the form,

\begin{equation}
D_\phi(r) = \ltp \frac{r}{\mrdiff} \rtp^{\beta-2} \;\;\; ,
\end{equation}
under the assumption of a power-law spectrum of density
inhomogeneities.  M13 provide expressions for \rdiff\
(see their Equation~7a) in two regimes:
  (i) $\mrdiff < \ell_0$, with $\ell_0$ the inner scale of 
  the inhomogeneities and;
  (ii) $\mrdiff \gg \ell_0$.
In the following\footnote{ 
The software accompanying this paper 
allows for alternate
choices for $\ell_0$ and $\beta$.},
we take $\ell_0 = \dlali$ 
and adopt a Kolmogorov spectrum with $\beta = 11/3$.

In the M13 formalism, \rdiff\ is a function of the 
Scattering Measure (SM) which is an integral over the 
squared amplitude of density fluctuations along the
sightline, $C^2_N(s)$.
For extragalactic calculations, one typically introduces
an effective Scattering Measure:

\begin{equation}
\msmeff = \int \frac{C^2_N (s)}{(1+z')^2} \, ds
\end{equation}
Following Equation~29 of M13, we estimate 
\begin{equation}
\msmeff \approx 5.6 \times 10^{16} \, {\rm m^{-17/3}} \;
  \ltp \frac{n_e}{10^{-2} \cm{-3}} \rtp^2 
  \ltp \frac{L_0}{10^{-3} \, \rm pc} \rtp^{-2/3} 
  \ltp \frac{\Delta L}{1 \, \rm kpc} \rtp
  \ltp \frac{2}{1 + z_{\rm DLA}} \rtp^{-2}
\end{equation}
with $\Delta L$ our adopted `size' for a DLA.
With the \smeff\ given above, we estimate
$\mrdiff = 2.2 \times 10^8 \, \rm m$ for
$\lambda_0 = 30$\,cm. 
This implies an essentially negligible
angular broadening of  $\approx 0.02$\,mas
for any source far behind the DLA.
We conclude that the ISM 
of galaxies intervening FRBs will insignificantly
broaden the angular size of any background source.

\subsection{Temporal Broadening}

The other significant effect from turbulent scattering is
that multi-path propagation through the medium leads to
temporal broadening of the burst.  This effect may yield
an FRB with much longer observed duration than the intrinsic event
and even preclude detection by current and planned
experiments \citep[e.g.][]{chawla17}.

Following M13, the temporal smearing may be related
to angular scattering,

\begin{figure}
	\includegraphics[width=\columnwidth]{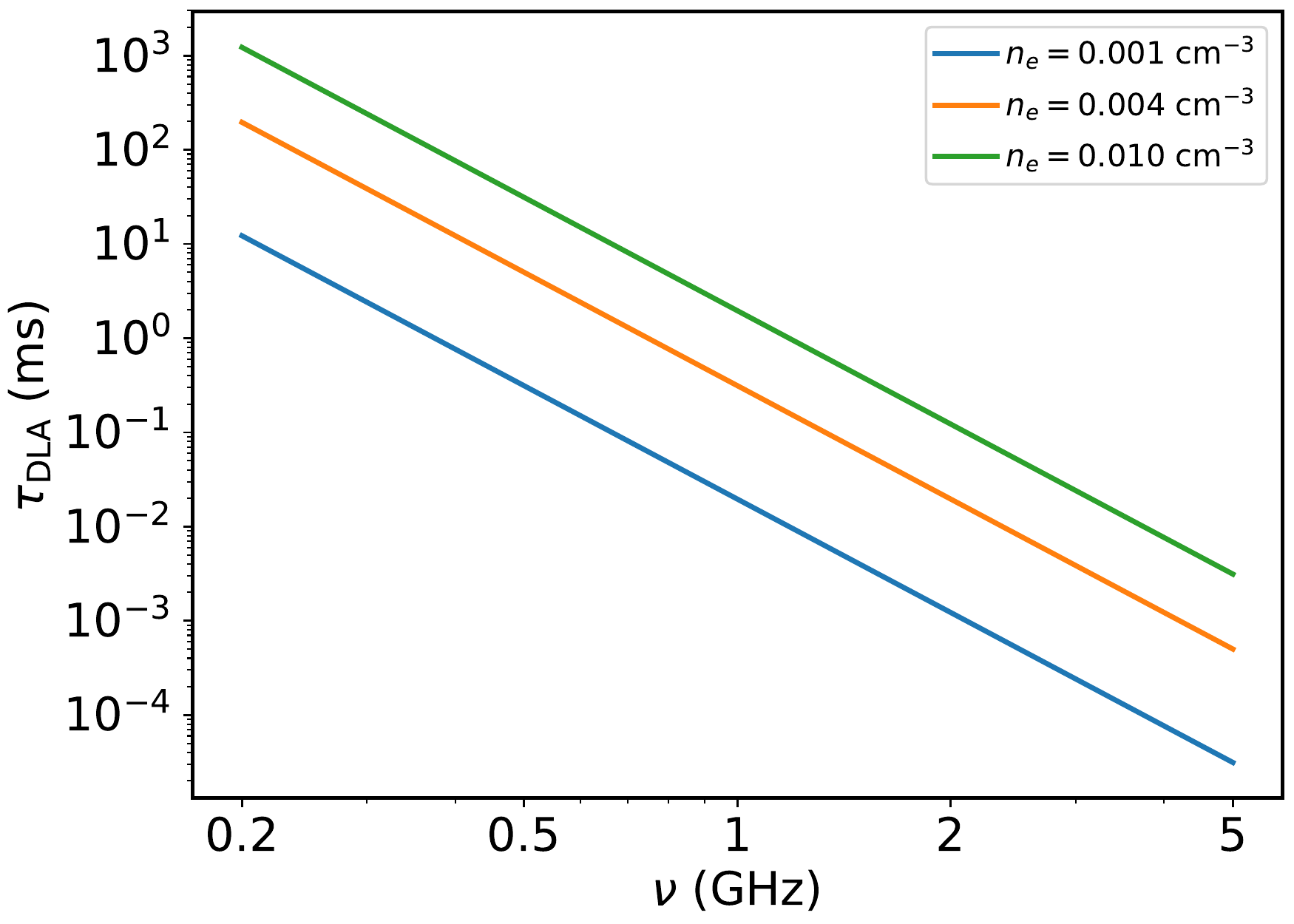}
    \caption{Estimates of the temporal broadening of an 
    FRB by a single, fiducial DLA \dlatau\ as a function
    of observed frequency.  
    The DLA has $l_0 = 1$\,AU, $\Delta L = 1$\,kpc, $L_0 = 0.001$\,pc,
    $z_{\rm DLA} = 1$, and the source is assumed to be at $z_{\rm FRB}=2$.
    The figure shows values for 
    electron densities spanning best estimations \citep{neeleman+15}.
    }
    \label{fig:tau_vs_nu}
\end{figure}

\begin{equation}
\mtscatt = \frac{D_L D_S \mascatt^2}{c D_{LS} (1+z_L)} \;\;\; ,
\end{equation}
The $\mascatt^2$ term gives a very steep frequency dependence,
i.e. $\mtscatt \propto \nu^{-\gamma}$ with $\gamma \approx 4-5$,
and has an electron density dependence of 
$\mtscatt \propto n_e^2$. 
This is illustrated in Figure~\ref{fig:tau_vs_nu} for a fiducial
DLA with electron densities ranging about our adopted,
characteristic value.  For $n_e < 10^{-3} \cm{-3}$,
$\mdlatau < 1$\,ms even for $\nu = 400$\,MHz and the effects on
the FRB population are likely negligible.  Larger $n_e$ values,
however, may have observational consequences.
For $n_e = 10^{-2} \cm{-3}$ and $\nu = 1$\,GHz, 
\tscatt\ exceeds several ms 
which is comparable to the median width 
for the current set of observed FRBs \citep{frbcat}.
Events with $\mtscatt \gg 5$\,ms will have 
reduced signal-to-noise ($S/N \propto \mtscatt^{1/2}$)
and those exceeding 100\,ms will not trigger most, current
FRB search algorithms.


We further emphasize that \tscatt\ has an explicit
$(1+z)^{-1}$ dependence and \ascatt\ 
has the same factor implying $\mtscatt \propto (1+z)^{-3}$.
Given the incidence for DLAs remains small for $z_L < 1$,
their impact on broadening of the FRB signal is small.
From our analysis, we may estimate the fraction of FRB
sightlines originating at redshift \zfrb\ that will
be broadened by at least $\tau_{\rm min}$.
We performed a Monte Carlo simulation of 10$^7$
sightlines for a series of \zfrb\ values, 
randomly inserting DLAs from a Poisson distribution 
with mean \avgn.
Each DLA is then assigned a random \nhi\ value drawn
from $h(\mnhi)$ to estimate $\Delta L = \mnhi / n_{\rm HI}$
with $n_{\rm HI} = 0.1 \cm{-3}$, and a random electron
density.
For the latter, we have assumed a Gaussian distribution
in $\log n_e$ centered on $\log (n_e/\cm{-3}) = -2.6$
with a (0.5\,dex)$^2$ variance.
For sightlines with multiple DLAs, we add \tscatt\ in 
quadrature.  

\begin{figure}
	\includegraphics[width=\columnwidth]{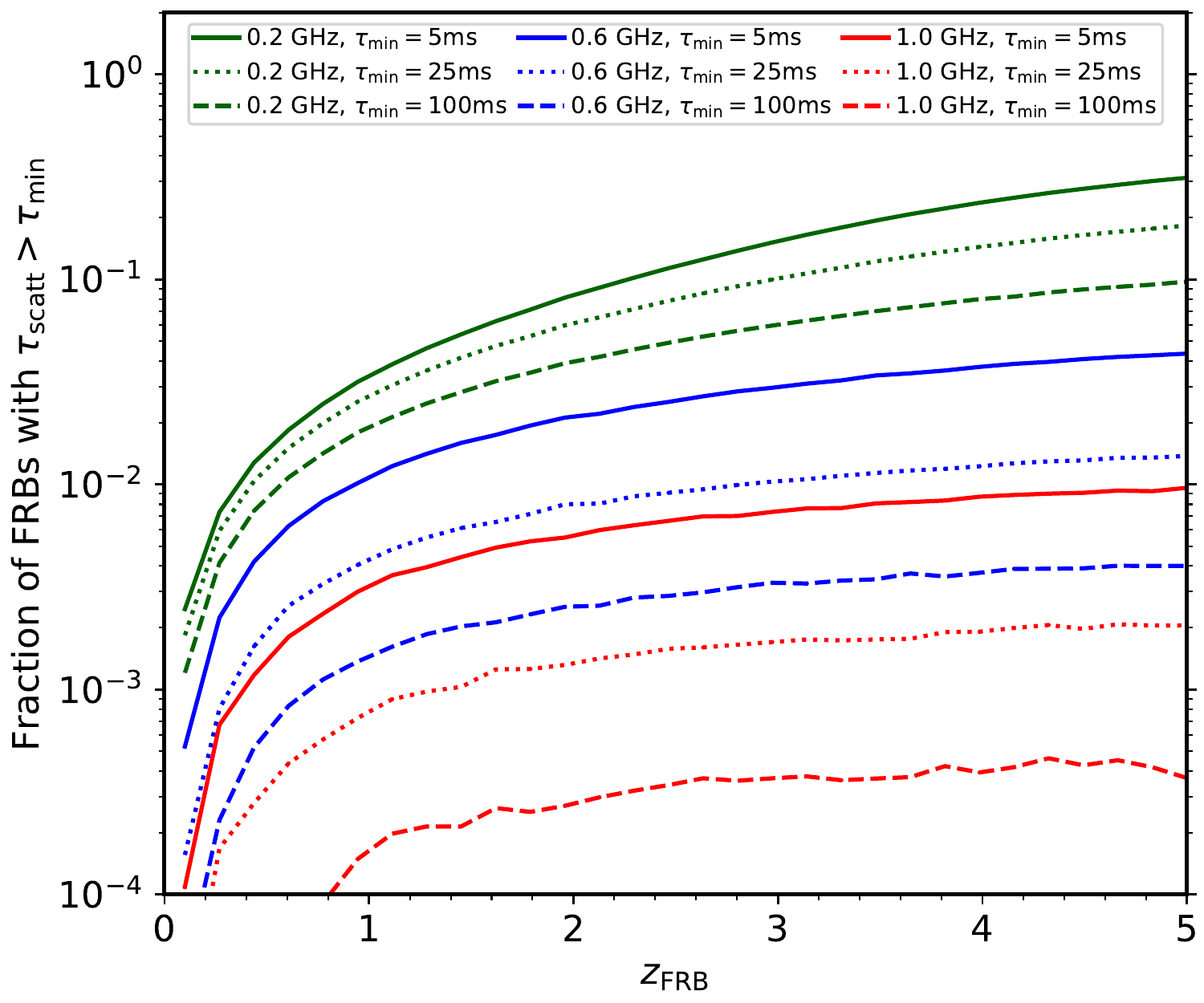}
    \caption{Fraction of sightlines for a population of FRBs
    at \zfrb\ predicted to experience $\mtscatt > \mtmin$
    for several assumed frequencies. The results are derived
    from a Monte Carlo simulation with random placement 
    of the redshifts, \nhi\ values, and electron densities
    [see the text for further details].  
    Even with $\nu = 600$\,MHz, the fraction of sightlines
    that are broadened beyond 25\,ms is only approximately 0.1\%.  
    We conclude that intersecting galaxies will have
    a nearly negligible influence on the observed distribution
    of FRBs.
    } 
    \label{fig:tau_lim}
\end{figure}

Figure~\ref{fig:tau_lim} shows
the results for several observed frequencies
and \tmin\ values.  Experiments with $\nu  = 1$\,GHz
are predicted to have only $\approx 0.1\%$ of their
sightlines broadened beyond 5\,ms.
We conclude that the ISM of intervening galaxies has 
negligible impact on any experiment with $\nu > 1$\,GHz
(e.g.\ REALFAST).  For $\nu = 600$\,MHz (e.g.\ CHIME),
the effects are modest but potentially impactful.
Approximately 1\%\ of the sightlines for FRBs at $\mzfrb > 1$
are broadened by greater than 5\,ms and 
$\approx 10^{-3}$ have $\mdlatau > 100$\,ms.
This implies a small but not entirely negligible
impact on the population of FRBs discovered
at low frequencies.
Experiments at even lower frequencies 
(e.g. LOFAR) will be affected, if the ISM is as modeled
here.  One can, of course, invert the situation and 
place constraints on ISM properties by surveying FRBs
at a range of frequency.


\section{Concluding Remarks}
In this manuscript, we have utilized surveys of DLAs
to infer the astrophysical impacts of the ISM in 
intervening galaxies on FRB events.  The principal
conclusion is that intervening galaxies will contribute
minimally to the integrated DM of an FRB.  For a DLA
at redshift \zdla, the average contribution from 
its neutral ISM is $\mdmnm =  \mdmdla (1+\mzdla)^{-1}$
as estimated from the \ion{H}{I} column density and
$n_e/n_{\rm H}$ distribution functions.  Given the
low incidence of intervening galaxies, the average
contribution along the sightline to an FRB is less
than $10^{-1} \mdmunits$.
From  $\N{Al^{++}}$ measurements in DLAs, we
estimate that the warm ionized component within these
galaxies typically contributes $\mdmwim < 20 \mdmunits$. 
Similarly, the angular and temporal broadening of
FRBs by the ISM of intervening galaxies is nearly
negligible for all but the lowest frequency experiments.

While our results indicate that FRB analysis will typically
not be sensitive to the ISM of galaxies, there are a few
notable exceptions.  With a large enough sample across
the sky, one may probe the ISM of our nearest neighbors
-- M31 and the Magellanic Clouds -- which span many
hundreds of square degrees.  
Even without precise
FRB localizations, one may attribute excess DM and/or 
temporal smearing from these galaxies.

It is also possible that FRBs frequently occur within
the ISM of distant, host galaxies and perhaps within 
regions of recent/ongoing star-formation.  
One has gained insight into the properties of such
environments from spectroscopy of long-duration 
gamma ray burst afterglows \citep[e.g.][]{pcb06,fjp+09}.
These data reveal highly elevated column densities of 
neutral hydrogen and associated metals but less evidence
for large column densities of ionized gas.  
There are, however, notable examples of highly 
saturated high-ion transitions 
\citep[e.g.\ \ion{Si}{IV}, \ion{C}{IV}][]{Mir03}
that indicate large electron column densities and
correspondingly large DM values.
By assessing the DM contribution to FRBs from
gas within their host environments, one will
gain new insight into the ISM of distant galaxies.

Throughout this manuscript, we have carefully distinguished
between gas within intervening galaxies (i.e.\ the ISM)
from gas surrounding such galaxies (a.k.a. halo gas or the CGM).
The latter exhibits a much large cross-section to absorption
and is likely the dominant baryonic component of galaxies
\citep[e.g.][]{werk+14}.  In a companion manuscript 
(Prochaska \& Zheng 2017, in prep.), we consider the astrophysical
impacts of halo gas on FRBs as inferred from surveys of our Galaxy
and the CGM of distant galaxies in absorption.

\section*{Acknowledgements}

We acknowledge valuable input from scientists
C. Law, J. Lazio, K. Bannister, and S. Tendulkar.




\bibliographystyle{mnras}
\bibliography{allrefs} 



\begin{table*}
\centering
\begin{minipage}{170mm} 
\caption{\nhi\ MEASUREMENTS\label{tab:param}}
\begin{tabular}{lccccccc}
\hline 
Measure & Param. & Unit & Value & 16th & 84th 
& 0.5th & 99.5th\\ 
\hline 
$h(\mnhi)$& $N_d$ & & $21.551$\\ 
       & $\alpha_3$ & & $-2.055$\\ 
       & $\alpha_4$ & & $-6.000$\\ 
$\ell_{\rm DLA}(z)$& $A$ & & $0.236$& $0.214$ & $0.255$& $0.186$ & $0.305$\\ 
            & $B$ & & $0.168$& $0.150$ & $0.177$& $0.132$ & $0.214$\\ 
            & $C$ & & $2.869$& $2.717$ & $3.020$& $2.465$ & $3.323$\\ 
\dmodla$^a$ && pc \, $\cm{-3}$ & 0.18& 0.5\,dex & 0.5\,dex& 1.0\,dex & 1.0\,dex\\ 
$\mfavgdm{1}$ && pc \, $\cm{-3}$ & 0.004& $0.000$ & $0.000$& $0.000$ & $0.183$\\ 
$\mfavgdm{2}$ && pc \, $\cm{-3}$ & 0.010& $0.000$ & $0.000$& $0.000$ & $0.218$\\ 
\dlatau$^b$ && ms & 0.31\\ 
\hline 
\end{tabular} 
\end{minipage} 
{$^a$}Rest-frame value.  Error is dominated by uncertainty in $n_e$.\\ 
{$^b$}Assumes $\nu=1$GHz, $n_e = 4 \times 10^{-3} \cm{-3}$, $z_{\rm DLA} = 1$, $z_{\rm source} = 2$.\\ 
\end{table*}


\bsp	
\label{lastpage}
\end{document}